\documentclass[prd,aps,nofootinbib,floatfix,11pt]{revtex4}
\usepackage{amsmath,graphicx,epsfig,amssymb,dsfont,mathtools}
\usepackage[usenames]{color}
\usepackage{ulem} 
\usepackage{bigstrut}
\usepackage{slashed}
\usepackage{multirow}
\usepackage{subfigure}
\usepackage{amsmath}
\usepackage{booktabs}

\allowdisplaybreaks

\begin{document}
	\title{Study of $\Lambda^0_b \to \Lambda D$ decays with the rescattering mechanism}
	\author{Na Li}
	\email{d202580107@hust.edu.cn}
	\affiliation{School of physics, Huazhong University of Science and Technology, Wuhan 430074, China }
	\author{Hui-Qiang Shang}
	\email{she@hust.edu.cn}
	\affiliation{School of physics, Huazhong University of Science and Technology, Wuhan 430074, China }
	\author{Qin Qin}
	\email{qqin@hust.edu.cn}
	\affiliation{School of physics, Huazhong University of Science and Technology, Wuhan 430074, China }

	\begin{abstract}
		We employ the final-state rescattering mechanism to systematically investigate the non-leptonic decays $\Lambda_b^0 \to \Lambda D$ (where $D = D^0$, $\overline{D}^0$, $D_+$, $D_-$), which can help improve the experimental precision of the $CP$-violating phase angle $\gamma$. The framework integrates short-distance factorizable amplitudes with long-distance non-factorizable contributions arising from hadronic triangle rescattering diagrams.
		We find that the final-state rescattering contributions are essential, not only because they provide the strong phases necessary for $CP$ violation, but also because they enhance the $\overline{D}^0 \Lambda$ branching fraction by two orders of magnitude. Numerically, we calculate the branching ratios of the channels with different partial waves, as well as the corresponding $CP$-violating observables. In particular, the direct $CP$ asymmetries
		$
		a_{CP}^{\mathrm{dir}}(D_+) = 0.28^{+0.07}_{-0.11}$
		and
		$
		a_{CP}^{\mathrm{dir}}(D_-) = -0.14^{+0.06}_{-0.04}$
		are found to be very significant. These results provide a reliable theoretical benchmark and can be tested against future measurements from the LHCb experiment.
	\end{abstract}
	
	\maketitle

	\section{Introduction}
	Exploring the origin of $C\!P$ violation is a crucial step in understanding the matter-antimatter asymmetry of the universe. According to the Sakharov conditions~\cite{Sakharov:1967dj}, $C\!P$ violation is an essential prerequisite for the generation of baryon asymmetry (the imbalance between matter and antimatter) during the evolution of the early universe. In the Standard Model (SM) of particle physics, $C\!P$ violation originates from a complex phase introduced by the mixing of three generations of quarks, which is parameterized by the Cabibbo-Kobayashi-Maskawa (CKM) matrix~\cite{Cabibbo:1965zzb,Kobayashi:1973fv}. This mechanism has been experimentally confirmed in the decays of $K$, $D$, $B$, and $B_s$ mesons~\cite{Christenson:1964fg,LHCb:2019hro,BaBar:2001ags,Belle:2001zzw,LHCb:2013syl,ParticleDataGroup:2022pth}. However, the $C\!P$ violation predicted by the SM is far too small to account for the observed baryon asymmetry of the universe. This significant discrepancy suggests the existence of $C\!P$ violation sources beyond the SM~\cite{Planck:2015fie}. Compared to mesons, the study of $C\!P$ violation in baryon systems holds unique value. First, because visible matter in the universe is primarily composed of baryons, investigating $C\!P$ violation in these systems links directly to cosmological observations. Furthermore, the richer helicity structure inherent in baryon decay amplitudes enables a larger set of observables, such as the decay asymmetry parameters $\alpha$, $\beta$, and $\gamma$. These observables open up new avenues for studies of quantum chromodynamics (QCD) and precision tests of the SM, while also serving as sensitive probes for new physics beyond the SM. With the accumulation of bottom baryon data at the Large Hadron Collider (LHC), a substantial number of experimental measurements on charmless non-leptonic decays of bottom baryons have been conducted~\cite{LHCb:2018fly,LHCb:2024iis,LHCb:2014yin,LHCb:2016rja,LHCb:2024yzj,LHCb:2025ray,LHCb:2019oke,LHCb:2018fpt,LHCb:2016yco}. Recently, the LHCb collaboration reported the first observation of $C\!P$ violation in baryon decays, specifically in $\Lambda^0_b \to p\pi^+K^-\pi^-$, with a $C\!P$ asymmetry of $(2.45 \pm 0.46 \pm 0.10)\%$~\cite{LHCb:2025ray}. This discovery not only confirms the existence of $C\!P$ violation in baryon systems but also opens up a new avenue for in-depth studies of the mechanism of $C\!P$ symmetry breaking in the baryon sector.
	
	Theoretically, non-leptonic decays of bottom baryons serve as an important avenue for testing the SM, exploring $C\!P$ violation mechanisms, and precisely determining CKM matrix elements. Currently, a variety of theoretical methods have been developed to study $\Lambda^0_b$ decays~\cite{Han:2024kgz,Lu:2009cm,Zhu:2018jet,Roy:2019cky,He:2015fsa,Geng:2016gul,Hsiao:2017tif,Ke:2007tg,Wei:2009np,Zhu:2016bra,Hsiao:2014mua,Khodjamirian:2011jp,Detmold:2015aaa,Feldmann:2011xf,He:2015fwa}. Among these, the $\Lambda^0_b \rightarrow \Lambda D$ decay channel is considered an ideal process for determining the CKM angle $\gamma$ because it is intrinsically free from penguin pollution~\cite{Giri:2001ju,Giri:2004wb,Zhu:2018jet,Zhang:2021sit}. Ref.~\cite{Zhang:2021sit} shows that the inclusion of the decay asymmetry parameter $\alpha$ as an additional observable can enhance the measurement sensitivity of $\gamma$ in these decays by up to 60\%. Previous theoretical efforts have evaluated the branching fractions of this process using various approaches, including the generalized factorization approach in Ref.~\cite{Giri:2001ju}, the improved phenomenological factorization method adopted by Zhu et al.~\cite{Zhu:2018jet}, and the bag model analysis by Geng et al.~\cite{Geng:2022osc}. However, for color-suppressed channels like $\Lambda^0_b \rightarrow \Lambda D$, branching fractions predicted solely by the naive factorization of short-distance operators tend to be significantly underestimated, and it is difficult to generate the sufficient strong phases required for $C\!P$ violation. Recently, a concurrent study has addressed this issue by evaluating the short-distance non-factorizable contributions within the perturbative QCD (PQCD) framework~\cite{Rui:2026ihu}. While their approach provides valuable insights from a short-distance perspective, developing a complementary theoretical framework that systematically incorporates long-distance non-perturbative contributions remains crucial. In this work, we introduce these long-distance contributions via final-state rescattering~\cite{Cheng:2004ru,Yu:2017zst,Jia:2024pyb,Duan:2024zjv,Shang:2026knt,Hu:2024uia,Li:2024fmg,Hu:2025pjg,Xing:2023kjk}. This mechanism decomposes the decay into two steps: first, the $\Lambda^0_b$ decays via the weak interaction into an intermediate hadronic pair (such as $\Lambda_c^+ K^{(*)}$ or $p D_s^{(*)}$), which is described by short-distance contributions under the factorization hypothesis. Subsequently, these intermediate hadrons undergo rescattering via single-particle exchange to form the final $\Lambda D$ state, a process visually represented by  single-particle exchange triangle diagrams. 
	
	Motivated by these considerations, we present a systematic study of the $\Lambda^0_b \to \Lambda D_\pm$ decay process, where $D_\pm = (D^0 \mp \bar{D}^0)/\sqrt{2}$ denote the $C\!P$ eigenstates. The remainder of this paper is organized as follows. In Section II, we establish the complete theoretical framework, decomposing the decay amplitude into short-distance contributions based on the factorization hypothesis and long-distance non-factorizable contributions based on the final-state rescattering mechanism. The short-distance contributions involve the $\Lambda^0_b \to \Lambda$ transition form factors and meson decay constants, while the long-distance contributions are precisely calculated via single-particle-exchange triangle diagrams. In Section III, we detail our numerical analysis, evaluating the long-distance effects using form factors from the covariant light-front quark model, combined with effective Lagrangians and strong coupling constants under $SU(3)$ flavor symmetry. Within the helicity amplitude formalism, we provide analytic definitions for the branching fractions, direct $C\!P$ asymmetry, partial-wave $C\!P$ asymmetry, and the decay asymmetry parameters ($\alpha$, $\beta$, $\gamma$) alongside their $C\!P$ conjugates. Subsequently, we present and discuss our numerical results in detail, comparing our branching fraction predictions with those of other theoretical models , and providing estimates for the observable final state branching fractions via hadronic $D$ meson decays. This study aims to fill the gap in the systematic understanding of $\Lambda^0_b \to \Lambda D$ decay dynamics, providing a reliable theoretical basis for precision tests of the SM, the measurement of the CKM angle $\gamma$, and the search for new physics signals in the baryon sector.
	
	\section{Theoretical framework}
	\label{Theoretical framework}
	In this study, we perform a systematic analysis of the non-leptonic weak decays $\Lambda^0_b \to \Lambda D_\pm$, where $D_\pm \equiv (D^0 \mp \bar{D}^0)/\sqrt{2}$ represents the $C\!P$ eigenstates of the $D$ meson. At the quark level, these processes are governed by two distinct tree-level transitions: $b \to c\bar{u}s$, corresponding to the $\Lambda^0_b \to \Lambda D^0$ mode, and $b \to u\bar{c}s$, corresponding to the $\Lambda^0_b \to \Lambda \bar{D}^0$ mode. The respective topological diagrams are illustrated in Fig.~\ref{fig:quark}. Since both transitions contribute to the same final state, their decay amplitudes interfere coherently. To provide a rigorous evaluation of the total amplitude, it is essential to incorporate both short-distance (SD) factorizable and long-distance (LD) non-factorizable contributions for each channel. The theoretical formalisms for these two types of contributions are detailed in the subsequent subsections.
	\subsection{Short-distance factorizable contributions}
	\label{sec:short_distance}
	
\begin{figure}[htbp]
		\centering
		\includegraphics[width=0.7\columnwidth]{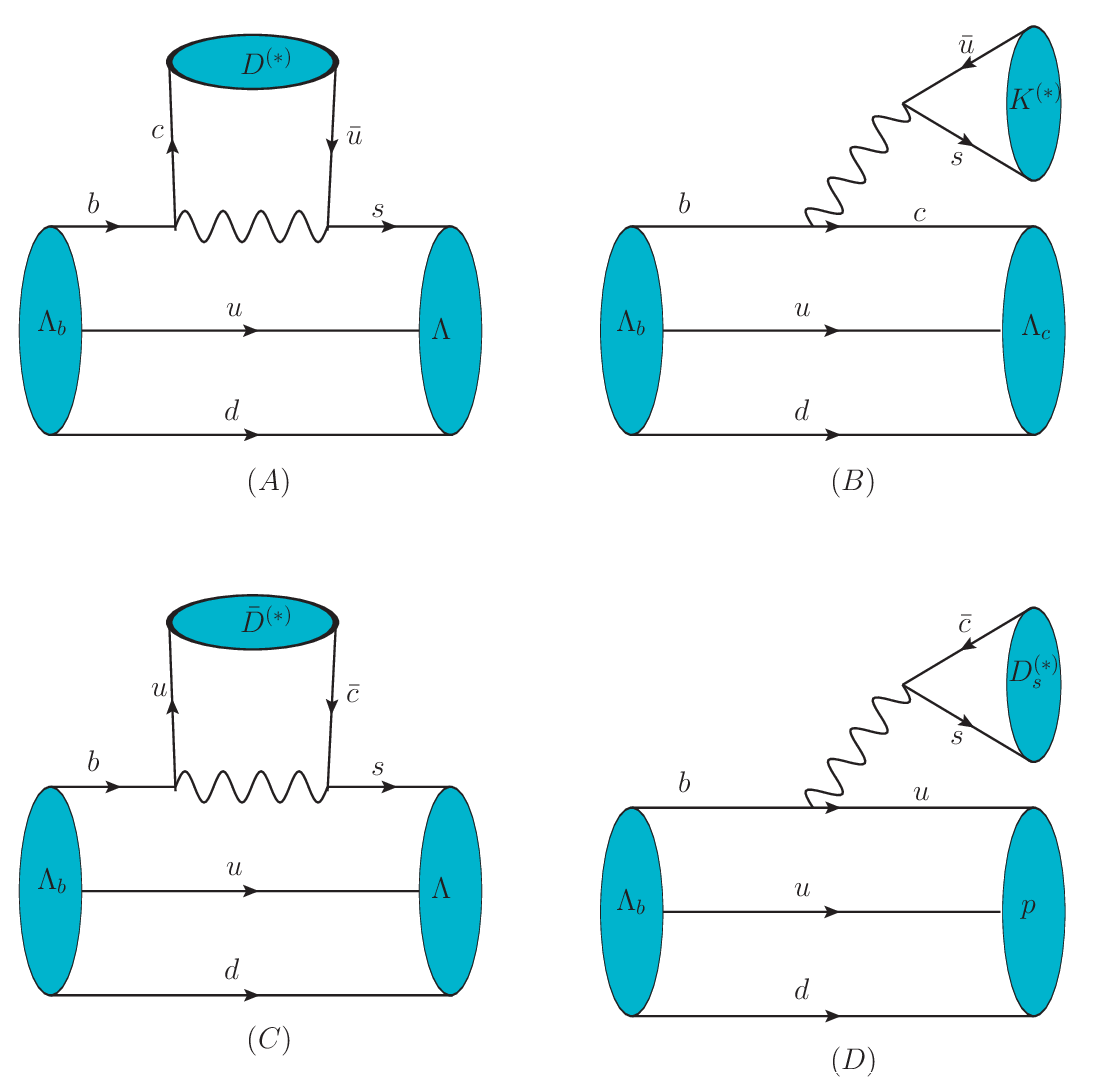}
		\caption{Quark-level tree diagrams for $\Lambda^0_b$ weak decays. Panels (A, B) and (C, D) correspond to the $b \rightarrow c\bar{u}s$ and $b \rightarrow u\bar{c}s$ transitions, respectively. Within each transition, the diagrams represent internal (A, C) and external (B, D) $W$-emission processes.}
		\label{fig:quark}
	\end{figure}

	In this section, we present the theoretical framework for evaluating the SD factorizable contributions, which form the dynamical foundation for estimating the LD non-factorizable effects via the final-state rescattering mechanism. The weak decay processes $\Lambda^0_b \to \mathcal{B}_1 M_1$, where the intermediate hadronic states denote $\mathcal{B}_1 M_1 = \Lambda_c^+ K^{-(*)}, \Lambda D^{0(*)}, p D_s^{-(*)}$, proceed through the factorizable $W$-emission diagrams as illustrated in Fig.~\ref{fig:quark}.
	
	The corresponding effective Hamiltonian is given by~\cite{Buchalla:1995vs,Lu:2009cm}:
	\begin{equation}
		\mathcal{H}_{\text{eff}} = \frac{G_F}{\sqrt{2}} V_{\text{CKM}} \left[ C_1(\mu) \mathcal{O}_1(\mu) + C_2(\mu) \mathcal{O}_2(\mu) \right] + \text{h.c.},
	\end{equation}
	where $G_F$ is the Fermi coupling constant, while $C_{1,2}(\mu)$ and $\mathcal{O}_{1,2}(\mu)$ are the Wilson coefficients and the local four-fermion operators, respectively, evaluated at the renormalization scale $\mu$. For the $b \to c\bar{u}s$ transition, for instance, the four-fermion operators are defined as $\mathcal{O}_1 = (\bar{c}_\alpha b_\beta)_{V-A} (\bar{s}_\beta u_\alpha)_{V-A}$ and $\mathcal{O}_2 = (\bar{c}_\alpha b_\alpha)_{V-A} (\bar{s}_\beta u_\beta)_{V-A}$, with $\alpha, \beta$ being the color indices. Under the naive factorization hypothesis, the weak transition matrix element $\langle \mathcal{B}_1 M_1 | \mathcal{H}_{\text{eff}} | \Lambda^0_b \rangle$ can be factorized as
	\begin{equation}
		\langle \mathcal{B}_1 M_1 | \mathcal{H}_{\text{eff}} | \Lambda^0_b \rangle_{\text{SD}} =
		\frac{G_F}{\sqrt{2}} V_{cb} V_{us}^* a_{1(2)}(\mu)
		\langle \mathcal{B}_1 | \bar{c} \gamma^\mu (1 - \gamma_5) b | \Lambda^0_b \rangle
		\langle M_1 | \bar{s} \gamma_\mu (1 - \gamma_5) u | 0 \rangle,
	\end{equation}
	where the effective Wilson coefficients are given by $a_{1(2)} = C_{1(2)} + C_{2(1)}/N_c$, with $N_c$ being the number of colors, evaluated at the scale $\mu = m_b$.
	
	The baryonic transition matrix element $\langle \mathcal{B}_1 | (\bar{c}b)_{V-A} | \Lambda^0_b \rangle$ is parametrized in terms of six form factors~\cite{Zhu:2018jet}:
\begin{equation}
	\label{eq:form_factors}
	\begin{aligned}[b]
		&\langle \mathcal{B}_{1}(p_{2}, s_{2}) | {(\bar{c}b)}_{V-A} | \Lambda^0_b(p_{i}, s_{i}) \rangle \\
		&\quad = \bar{u}(p_{2}, s_{2}) \left[ \gamma_{\mu} f_{1}(q^2) + i\sigma_{\mu\nu}\frac{q^{\nu}}{m_{\Lambda^0_b}} f_{2}(q^2) + \frac{q_{\mu}}{m_{\Lambda^0_b}} f_{3}(q^2) \right] u(p_{i}, s_{i}) \\
		&\quad- \bar{u}(p_{2}, s_{2}) \left[ \gamma_{\mu} g_{1}(q^2) + i\sigma_{\mu\nu}\frac{q^{\nu}}{m_{\Lambda^0_b}} g_{2}(q^2) + \frac{q_{\mu}}{m_{\Lambda^0_b}} g_{3}(q^2) \right] \gamma_{5} u(p_{i}, s_{i}) \,,
	\end{aligned}
\end{equation}
	where $q = p_i - p_2$ is the momentum transfer, while $u(p_i, s_i)$ and $\bar{u}(p_2, s_2)$ denote the Dirac spinors for the initial $\Lambda^0_b$ and the final state baryon $\mathcal{B}_1$ ($\Lambda_c^+, \Lambda$, or $p$), respectively. The form factors $f_j(q^2)$ and $g_j(q^2)$ ($j=1,2,3$) are determined via non-perturbative methods, and their numerical values are summarized in Table~\ref{tab:form_factors_large}.
	
	The matrix elements for meson creation from the vacuum, $\langle M_1 | (\bar{q}_1 q_2)_{V-A} | 0 \rangle$, are parametrized by the corresponding meson decay constants:
	\begin{align}
		\langle P(p_1) | (\bar{q}_1 q_2)_{V-A} | 0 \rangle &= i f_P p_1^\mu, \nonumber \\
		\langle V(p_1) | (\bar{q}_1 q_2)_{V-A} | 0 \rangle &= m_V f_V \varepsilon^{* \mu}(p_1),
	\end{align}
	where $P$ and $V$ denote the pseudoscalar ($D_{(s)}$ or $K$) and vector ($D_{(s)}^*$ or $K^*$) mesons, respectively. Here, $p_1$ is the four-momentum of the meson, and $\varepsilon^{* \mu}(p_1)$ represents the polarization vector of the vector meson.
	
	Consequently, the invariant amplitudes $\mathcal{M}_{\text{SD}}(\Lambda^0_b \to \mathcal{B}_1 M_1)$ can be expressed as
	\begin{align}
		\mathcal{M}\left(\Lambda^0_b(p_i) \to \mathcal{B}_1(p_2) P(p_1)\right) &= i \bar{u}(p_2) \left[ A_1 + A_2 \gamma_5 \right] u(p_i), \label{eq:amp_P} \\
		\mathcal{M}\left(\Lambda^0_b(p_i) \to \mathcal{B}_1(p_2) V(p_1)\right) &= \bar{u}(p_2) \left[ B_1 \gamma_\alpha \gamma_5 + B_2 \frac{p_{2\alpha}}{m_{\Lambda^0_b}} \gamma_5 + B_3 \gamma_\alpha + B_4 \frac{p_{2\alpha}}{m_{\Lambda^0_b}} \right] \varepsilon^{* \alpha}(p_1) u(p_i), \label{eq:amp_V}
	\end{align}
	where the invariant amplitude components $A_j$ and $B_j$ are explicitly derived under the factorization framework as
	\begin{align}
		A_1 &= \lambda_{1(2)} f_P \left[ (m_{\Lambda^0_b} - m_{\mathcal{B}_1}) f_1(p_1^2) + \frac{p_1^2}{m_{\Lambda^0_b}} f_3(p_1^2) \right], \nonumber \\
		A_2 &= \lambda_{1(2)} f_P \left[ (m_{\Lambda^0_b} + m_{\mathcal{B}_1}) g_1(p_1^2) - \frac{p_1^2}{m_{\Lambda^0_b}} g_3(p_1^2) \right], \nonumber \\
		B_1 &= -\lambda_{1(2)} m_V f_V \left[ g_1(p_1^2) + \frac{m_{\Lambda^0_b} - m_{\mathcal{B}_1}}{m_{\Lambda^0_b}} g_2(p_1^2) \right], \nonumber \\
		B_2 &= -2 \lambda_{1(2)} m_V f_V g_2(p_1^2), \nonumber \\
		B_3 &= \lambda_{1(2)} m_V f_V \left[ f_1(p_1^2) - \frac{m_{\Lambda^0_b} + m_{\mathcal{B}_1}}{m_{\Lambda^0_b}} f_2(p_1^2) \right], \nonumber \\
		B_4 &= 2 \lambda_{1(2)} m_V f_V f_2(p_1^2),
		\label{eq:invariant_amplitudes}
	\end{align}
	with the coefficient factor defined as $\lambda_{1(2)} = \frac{G_F}{\sqrt{2}} V_{cb} V_{us}^* a_{1(2)}(\mu)$.
	\subsection{Long-distance non-factorizable contributions from final-state rescattering}
	\label{sec:long_distance}
	
\begin{figure}[htbp]
		\centering
		\includegraphics[width=0.99\columnwidth]{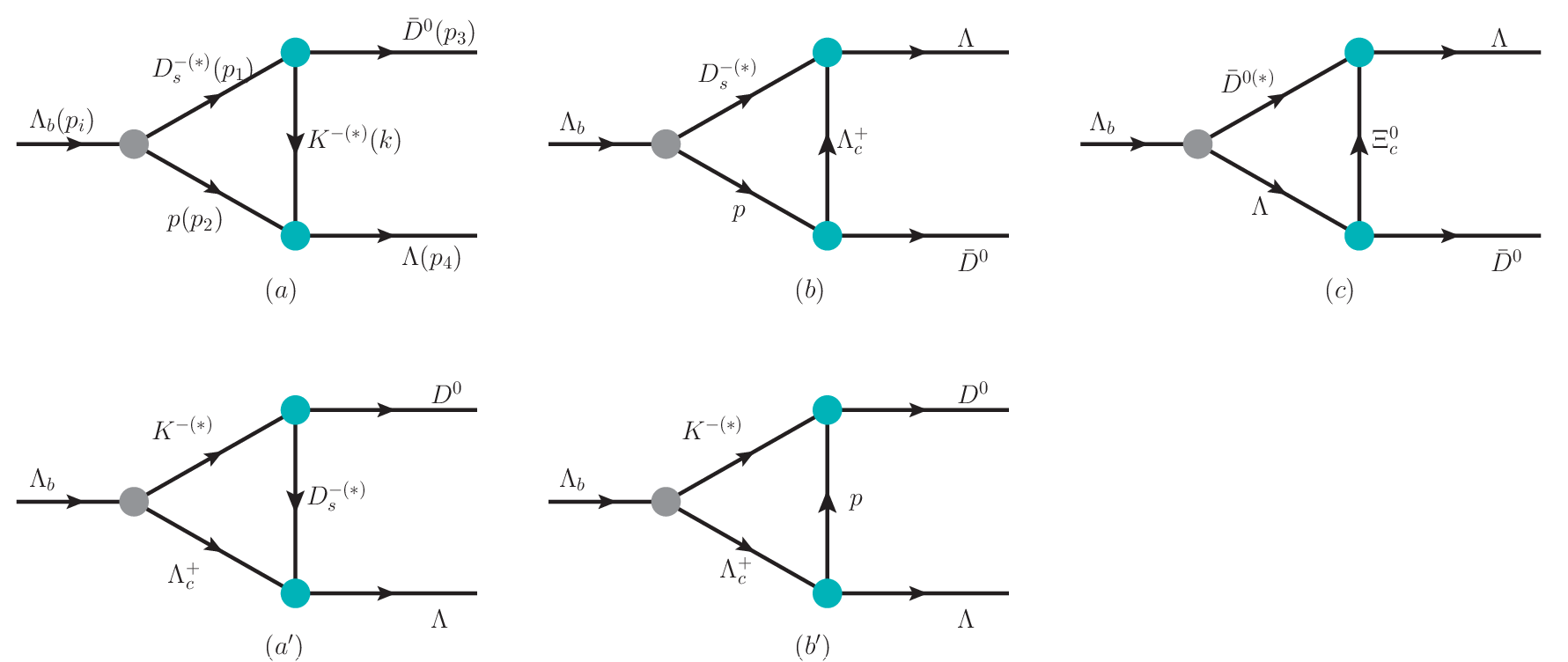}
		\caption{Hadronic triangle diagrams and kinematic notations for the final-state rescattering contributions to $\Lambda^0_b \to \Lambda D^0(\bar D^0)$ decays via one-particle exchange.}
		\label{fig:LambdaD}
	\end{figure}

	Although long-distance  non-factorizable contributions are physically significant in heavy baryon decays, they are notoriously difficult to evaluate from first principles. In this section, we employ the final-state rescattering  mechanism to systematically estimate these contributions, treating the hadronic interactions within the framework of effective Lagrangians. By inserting a complete set of intermediate hadronic states, the total matrix element for the decay into a specific flavor eigenstate, such as $\Lambda^0_b \to \Lambda D^0$, can be factorized into a sum over the intermediate states. This sum involves the product of the short-distance weak transition matrix element and the long-distance strong rescattering matrix element:
	\begin{equation}
		\langle \Lambda D^0 | \mathcal{H}_{\text{eff}} | \Lambda^0_b \rangle = \sum_{\mathcal{B}_1 M_1} \langle \Lambda D^0 | \mathcal{L} | \mathcal{B}_1 M_1 \rangle \langle \mathcal{B}_1 M_1 | \mathcal{H}_{\text{eff}} | \Lambda^0_b \rangle_{\text{SD}},
	\end{equation}
	where the summation runs over all kinematically allowed intermediate hadronic states $\mathcal{B}_1 M_1$, such as $\Lambda_c^+ K^{-(*)}$, $p D_s^{-(*)}$, or $\Lambda D^{0(*)}$. The term $\langle \Lambda D^0 | \mathcal{L} | \mathcal{B}_1 M_1 \rangle$ denotes the strong transition matrix element governed by the effective Lagrangian $\mathcal{L}$. This rescattering process can be visually described by the reaction chain $\Lambda^0_b \to \mathcal{B}_1 M_1 \xrightarrow{M/\mathcal{B}} \Lambda D^0$, which corresponds to the hadronic triangle diagrams illustrated in Fig.~\ref{fig:LambdaD}. Within the one-particle exchange approximation, the exchanged hadronic state can be either a meson ($M = K^{-(*)}, D_s^{-(*)}$) or a baryon ($\mathcal{B} = p, \Lambda_c^+, \Xi_c^0$).
	
	The evaluation of these triangle loops requires inputs for both the weak and strong interaction vertices. The weak vertices correspond to the short-distance factorized processes shown in Fig.~\ref{fig:quark}, with their amplitudes explicitly derived in Sec.~\ref{sec:short_distance}. The strong vertices are determined by the effective Lagrangians. The complete set of effective Lagrangians employed in this work, alongside their corresponding coupling constants, is summarized in Appendix A. The corresponding amplitudes for the $\bar{D}^0$ channel can be obtained analogously.
	
	Since the strong coupling constants are conventionally extracted under on-shell conditions, whereas all three internal hadrons in the triangle diagrams are off-shell, it is necessary to introduce a phenomenological form factor $\mathcal{F}(\Lambda, m_k)$. This form factor accounts for the off-shell effects at the vertices and the finite physical size of the hadrons, while simultaneously serving as a UV cutoff to regularize the loop integrals. We adopt the following monopole parameterization~\cite{Duan:2024zjv,Shang:2026knt}:
	\begin{equation}
		\mathcal{F}(\Lambda, m_k) = \frac{\Lambda^4}{(k^2 - m_k^2)^2 + \Lambda^4},
	\end{equation}
	where $m_k$ and $k$ are the mass and four-momentum of the exchanged particle, respectively, and the cutoff parameter is constrained to $\Lambda = 1.0 \pm 0.1 \ \mathrm{GeV}$~\cite{Duan:2024zjv,Shang:2026knt}.
	
	Having specified the weak and strong vertices as well as the hadronic propagators, the analytic amplitudes for the triangle diagrams can be evaluated by performing the loop integrals. Taking the $\Lambda_b^0 \to \Lambda D^0$ channel as an illustrative example, the corresponding triangle diagrams are depicted in Fig.~\ref{fig:LambdaD}($a^{\prime}-b^{\prime}$). The integrated amplitudes read
	\begin{align}
		\mathcal{M}_{a^\prime}[K^{*-},\Lambda^+_c;D^-_s]
		&=g_{ D^-_s D^0 K^{*-}} {g_{\Lambda^+_c \Lambda D^-_s }} \int \frac{d^4k}{(2\pi)^4} \bar{u}(p_4,\lambda_4)\gamma_5 (\not\!p_2+m_2)(B_1\gamma_{\nu}\gamma_5+B_2\frac{p_{2\nu}}{{m_{\Lambda^0_b}}}\gamma_5 
		\nonumber\\&+{B}_3\gamma_{\nu}+B_4\frac{p_{2\nu}}{{m_{\Lambda^0_b}}})u(p_i,\lambda_i)\frac{-(-g^{\alpha\nu}+\frac{p_1^{\alpha} p_1^{\nu}}{m_1^2}){(p_3-k)}_{\alpha}{\mathcal{F}}}{(p_1^2-m_1^2+i\varepsilon)(p_2^2-m_2^2+i\varepsilon)(k^2-m_k^2+i\varepsilon)},\nonumber\\
		\mathcal{M}_{a^\prime}[K^{*-},\Lambda^+_c;D^{*-}_s]
		&=2ig_{ D^{*-}_s D^0 K^{*-}} \int \frac{d^4k}{(2\pi)^4}\bar{u}(p_4,\lambda_4)\left(-{g_{\Lambda^+_c \Lambda D^{*-}_s }}\gamma^{\mu}-\frac{if_{\Lambda^+_c \Lambda D^{*-}_s }}{m_2+m_4}\sigma^{\mu\nu}k_{\nu}\right){p_1^{\alpha}}\nonumber\\ 
		&(\not\!p_2 +m_2)\bigg(B_1\gamma_{\delta}\gamma_5 +B_2\frac{p_{2\delta}} {m_i}\gamma_5
		+{B}_3\gamma_{\delta} +B_4\frac{p_{2\delta}}{m_i}\bigg){u(p_i,\lambda_i)\varepsilon_{\alpha\beta\rho\sigma}{(k-p_3)}^{\rho}}\nonumber\\
		& \frac{(-g^{\delta\beta}+\frac{p_1^{\delta} p_1^{\beta}}{m_1^2})(-g^{\sigma\mu}+\frac{k^{\sigma} k^{\mu}}{m_k^2})\mathcal{F}}{(p_1^2-m_1^2+i\varepsilon)(p_2^2-m_2^2+i\varepsilon)(k^2-m_k^2+i\varepsilon)},\nonumber\\
		\mathcal{M}_{a^\prime}[ K^-,\Lambda^+_c;D^{*-}_s]
		&=g_{ D^{*-}_s D^0 K^{-}}\int\frac{d^4k}{(2\pi)^4}\bar{u}(p_4,\lambda_4)(-{g_{\Lambda^+_c \Lambda D^{*-}_s }}\gamma^{\mu}-\frac{if_{\Lambda^+_c \Lambda D^{*-}_s }}{m_2+m_4}\sigma^{\mu\nu}k_{\nu})(\not\!p_2+m_2)\nonumber\\
		&\frac{(A_1+A_2\gamma_5)u(p_i,\lambda_i)(-g_{\mu\alpha}+\frac{k_\mu k_\alpha}{m_k^2}){p_1}_\alpha{\mathcal{F}}}{(p_1^2-m_1^2+i\varepsilon)(p_2^2-m_2^2+i\varepsilon)(k^2-m_k^2+i\varepsilon)},\nonumber\\
		\mathcal{M}_{b^\prime}[K^-,\Lambda^+_c; p]
		&= {g_{\Lambda p K^-} g_{\Lambda_c^+ p D^0}}\int \frac{d^4k}{(2\pi)^4} \bar{u}(p_4,\lambda_4)  \gamma_5(\not\!k+m_k)\gamma_5(\not\!p_2+m_2) \nonumber\\
		&\quad \times \frac{(A_1+A_2\gamma_5)u(p_i,\lambda_i) \mathcal{F}}{(p_1^2-m_1^2+i\varepsilon)(p_2^2-m_2^2+i\varepsilon)(k^2-m_k^2+i\varepsilon)}, \nonumber\\
		\mathcal{M}_{b^\prime}[K^{*-},\Lambda^+_c;p]&= \int\frac{d^{4}k}{(2\pi)^{4}} {g_{\Lambda_c^+ p D^0}}\bar{u} 
		(p_4,\lambda_4)(-g_{\Lambda p K^{*-}}\gamma_{\mu}-\frac{if_{\Lambda p K^{*-}}}{m_2+m_4}\sigma_{\mu\nu}p_1^{\nu})(\not\!k+m_k)\gamma_5 \nonumber\\
		&(\not\!p_2+m_2)(B_1\gamma_{\beta}\gamma_5+B_2\frac{p_{2\beta}}{m_i}\gamma_5+{B}_3\gamma_{\beta}+B_4\frac{p_{2\beta}}{m_i})u(p_i,\lambda_i)(-g^{\beta \mu}+\frac{p_1^{\beta}p_1^\mu}{m_{1}^2})\nonumber\\
		&\frac{\mathcal{F}}{(p_1^2-m_1^2+i\varepsilon)(p_2^2-m_2^2+i\varepsilon)(k^2-m_k^2+i\varepsilon)},
	\end{align}
	where $m_{\Lambda_b^0}$, $m_1$, $m_2$, $m_4$, and $m_k$ denote the masses of the initial-state baryon, intermediate-state meson, intermediate-state baryon, final state baryon, and the exchanged particle, respectively, with the kinematic notations defined in Fig.~\ref{fig:LambdaD}. Note that $p_i$ refers to the four-momentum of the initial $\Lambda_b^0$. After incorporating all contributing terms, the total amplitude for the decay $\Lambda_b^0 \to \Lambda D^0$ reads
	\begin{align}
		\mathcal{M}(\Lambda^0_b \to \Lambda D^0) &= \mathcal{M}_{\text{SD}}(\Lambda^0_b \to \Lambda D^0) + \mathcal{M}_{a^\prime}[K^{*-},\Lambda^+_c; D^-_s] + \mathcal{M}_{a^\prime}[K^{*-},\Lambda^+_c; D^{*-}_s] \nonumber\\
		&\quad + \mathcal{M}_{a^\prime}[K^-,\Lambda^+_c; D^{*-}_s] + \mathcal{M}_{b^\prime}[K^-,\Lambda^+_c; p] + \mathcal{M}_{b^\prime}[K^{*-},\Lambda^+_c; p].
	\end{align}
	Here, $\mathcal{M}_{\text{SD}}$ denotes the short-distance factorizable contribution, while $\mathcal{M}_{a^\prime(b^\prime)}$ correspond to the long-distance triangle diagram corrections. The full amplitude for the decay channel $\Lambda^0_b \to \Lambda \bar{D}^0$ is analogous and is presented explicitly in Appendix A. Finally, the complete decay amplitude for $\Lambda^0_b \to \Lambda D_\pm$ is constructed via the coherent superposition:
	\begin{equation}
		\mathcal{M}(\Lambda^0_b \to \Lambda D_{\pm}) = \frac{1}{\sqrt{2}} \Big( \mathcal{M}(\Lambda^0_b \to \Lambda D^0) \mp \mathcal{M}(\Lambda^0_b \to \Lambda \bar{D}^0) \Big),
	\end{equation}
	where the positive and negative signs correspond to the $D_{+}$ and $D_{-}$ $C\!P$ eigenstates, respectively.
	\section{Numerical Analysis}
	\subsection{Input Parameters}
	For the CKM quark mixing matrix, we adopt the Wolfenstein parametrization up to $\mathcal{O}(\lambda_W^3)$. The corresponding parameter values are taken from the Particle Data Group (PDG)~\cite{ParticleDataGroup:2022pth} as $A = 0.823$, $\rho = 0.141$, $\eta = 0.349$, and $\lambda_W = 0.225$. The effective Wilson coefficients evaluated at the scale $\mu = m_b$ are taken from Ref.~\cite{Lu:2009cm}. Furthermore, the decay constants of the relevant mesons are summarized in Table~\ref{tab:decay_constants}.
	\begin{table}[htbp]
		\centering
		\renewcommand{\arraystretch}{1.3} 
		\caption{The decay constants of the relevant pseudo-scalar and vector mesons (in units of MeV)~\cite{FlavourLatticeAveragingGroupFLAG:2024oxs,Ball:2004rg,Lubicz:2017asp}.}
		\label{tab:decay_constants}
		\begin{tabular}{cccccc}
			\hline\hline
			$f_K$ & $f_D$ & $f_{D_s}$ & $f_{K^*}$ & $f_{D^*}$ & $f_{D_s^*}$ \\
			\hline
			$155.7 \pm 0.3$ & $212.0 \pm 0.7$ & $249.9 \pm 0.5$ & $217\pm 5$& $223.5\pm 8.4$ & $268.8\pm 6.6$ \\
			\hline\hline
		\end{tabular}
	\end{table}
	
	In the physical region, the $\Lambda_b^0 \to \Lambda$ transition form factors are parameterized using a three-parameter dipole-like form~\cite{Zhu:2018jet}:
	\begin{equation}
		F(q^2) = \frac{r_1}{1 - q^2/M_{\mathrm{fit}}^2} + \frac{r_2}{(1 - q^2/M_{\mathrm{fit}}^2)^2},
	\end{equation}
	where $F(q^2)$ generically represents the form factors $f_{1,2,3}(q^2)$ and $g_{1,2,3}(q^2)$. The corresponding parameters $r_1$, $r_2$, and the pole mass $M_{\mathrm{fit}}$ are listed in Table~\ref{tab:form_factors_large}.
	
	\begin{table}[htbp]
		\centering
		\renewcommand{\arraystretch}{1.2}
		\setlength{\tabcolsep}{6pt}
		\caption{Interpolation parameters for the $\Lambda_b^0 \to \Lambda$ form factors defined in Eq.~\eqref{eq:form_factors}. The parameters $r_1$ and $r_2$ are dimensionless, and the fit mass is universally set to $M_{\mathrm{fit}} = 6.2$~GeV~\cite{Zhu:2018jet} for all form factors.}
		\begin{tabular}{c c c c c c c}
			\hline\hline
			& $f_1$ & $f_2$ & $f_3$ & $g_1$ & $g_2$ & $g_3$ \\
			\hline
			$F(0)$     & 0.131 & -0.048 & -0.027 & 0.132 & -0.023 & -0.052 \\
			$r_1$      & -0.091 & 0.051 & 0.028 & -0.092 & 0.026 & 0.053 \\
			$r_2$      & 0.222 & -0.098 & -0.055 & 0.224 & -0.050 & -0.105 \\
			\hline\hline
		\end{tabular}
		\label{tab:form_factors_large}
	\end{table}
	
	The evaluation of the rescattering amplitudes relies on the effective Lagrangian approach~\cite{Cheng:2004ru,Aliev:2006xr,Aliev:2009ei,Yu:2017zst,Duan:2024zjv,Xiao:2019mvs}. Within the framework of $SU(3)$ flavor symmetry, the strong coupling constants governing the various hadronic vertices are interrelated. Their explicit values and detailed derivations are summarized in Appendix A.
	
	\subsection{Definitions of the Observables}
	\label{sec:asymmetry_parameters}
	
	In this section, the branching fractions and decay asymmetry parameters for the two-body non-leptonic decays $\Lambda_b^0 \to \Lambda D_\pm$ are derived within the helicity amplitude formalism. The decay width is obtained by summing the squared helicity amplitudes over all possible spin states of the final state particles and averaging over the initial spin states:
	\begin{equation}
		\label{eq:helicity_width}
		\Gamma(\Lambda_b^0 \to \Lambda D_\pm) = \frac{|\mathbf{p}_f|}{8\pi m_{\Lambda_b^0}^2} \frac{1}{2} \left( |\mathcal{H}_{+1/2}|^2 + |\mathcal{H}_{-1/2}|^2 \right),
	\end{equation}
	where $|\mathbf{p}_f| = \sqrt{[m_{\Lambda_b^0}^2 - (m_\Lambda + m_{D_\pm})^2][m_{\Lambda_b^0}^2 - (m_\Lambda - m_{D_\pm})^2]} / (2m_{\Lambda_b^0})$ is the magnitude of the three-momentum of the final state particles in the rest frame of the initial $\Lambda_b^0$ baryon. The factor of $1/2$ originates from the spin average over the initial unpolarized $\Lambda_b^0$ baryon.
	
	The helicity amplitudes are concisely defined as $\mathcal{H}_{+1/2} \equiv \mathcal{H}^{+1/2}_{+1/2,0}$ and $\mathcal{H}_{-1/2} \equiv \mathcal{H}^{-1/2}_{-1/2,0}$, where the superscript $\lambda_i = \pm 1/2$ labels the helicity of the initial-state baryon, while the subscripts $\lambda_f = \pm 1/2$ and $0$ denote the helicities of the final state baryon and meson, respectively. Due to the conservation of angular momentum, other helicity combinations vanish identically in the $\Lambda_b^0 \to \Lambda D_\pm$ decay; thus, the summation naturally reduces to the two terms presented in Eq.~\eqref{eq:helicity_width}. Consequently, the branching fraction is obtained by dividing the partial decay width by the total decay width of the $\Lambda_b^0$ baryon.
	
	The total direct $C\!P$ asymmetry is formulated as:
	\begin{equation}
		a_{CP}^{\text{dir}} = \frac{\Gamma - \overline{\Gamma}}{\Gamma + \overline{\Gamma}} = \frac{|S|^2 - |\overline{S}|^2 + |P|^2 - |\overline{P}|^2}{|S|^2 + |\overline{S}|^2 + |P|^2 + |\overline{P}|^2},
		\label{eq:cp_asym_total}
	\end{equation}
	where $\overline{\Gamma}$ denotes the decay width of the corresponding CP-conjugated process. The quantities $S$ and $P$ are the $S$-wave and $P$-wave parity-violating and parity-conserving decay amplitudes, respectively, while $\overline{S}$ and $\overline{P}$ represent their CP-conjugated counterparts. These partial-wave amplitudes are related to the helicity amplitudes via:
	\begin{equation}
		S = \frac{1}{\sqrt{2}} \big( \mathcal{H}_{+\frac{1}{2}} + \mathcal{H}_{-\frac{1}{2}} \big), \quad
		P = \frac{1}{\sqrt{2}} \big( \mathcal{H}_{+\frac{1}{2}} - \mathcal{H}_{-\frac{1}{2}} \big).
		\label{eq:partial_wave_amplitudes}
	\end{equation}
	Analogously, the isolated $C\!P$ asymmetries for the $S$-wave and $P$-wave sectors are defined as~\cite{Roy:2019cky}:
	\begin{equation}
		a_{CP}^S = \frac{|S|^2 - |\overline{S}|^2}{|S|^2 + |\overline{S}|^2}, \quad
		a_{CP}^P = \frac{|P|^2 - |\overline{P}|^2}{|P|^2 + |\overline{P}|^2}.
		\label{eq:cp_asym_sp_wave}
	\end{equation}
	
	In addition to the decay rate and the direct $C\!P$ asymmetry, the Lee-Yang asymmetry parameters $\alpha$, $\beta$, and $\gamma$ are conventionally utilized to characterize the angular distributions and polarization effects in baryon decays. They are expressed in terms of the partial-wave amplitudes as~\cite{Duan:2024zjv,Chen:2019hqi}:
	\begin{equation}
		\alpha = \frac{2\mathrm{Re}(S^* P)}{|S|^2 + |P|^2}, \quad
		\beta = \frac{2\mathrm{Im}(S^* P)}{|S|^2 + |P|^2}, \quad
		\gamma = \frac{|S|^2 - |P|^2}{|S|^2 + |P|^2}.
		\label{eq:asym_params}
	\end{equation}
	The corresponding parameters $\bar{\alpha}$, $\bar{\beta}$, and $\bar{\gamma}$ for the anti-baryon decays are defined in exactly the same manner using $\overline{S}$ and $\overline{P}$. Subsequently, the averaged asymmetry parameters and their associated CP-violating observables are constructed as~\cite{Donoghue:1986hh}:
	\begin{equation}
		\begin{aligned}
			\langle \alpha \rangle &= \frac{\alpha - \bar{\alpha}}{2}, \quad &\langle \beta \rangle &= \frac{\beta - \bar{\beta}}{2}, \quad &\langle \gamma \rangle &= \frac{\gamma + \bar{\gamma}}{2}, \\
			a_{CP}^{\alpha} &= \frac{\alpha + \bar{\alpha}}{2}, \quad &a_{CP}^{\beta} &= \frac{\beta + \bar{\beta}}{2}, \quad &a_{CP}^{\gamma} &= \frac{\gamma - \bar{\gamma}}{2}.
		\end{aligned}
		\label{eq:averaged_asym}
	\end{equation}
	
	\subsection{Numerical results}
	\label{sec:discussion}
	
\begin{figure}[htbp]
		\centering
		\includegraphics[width=0.99\columnwidth]{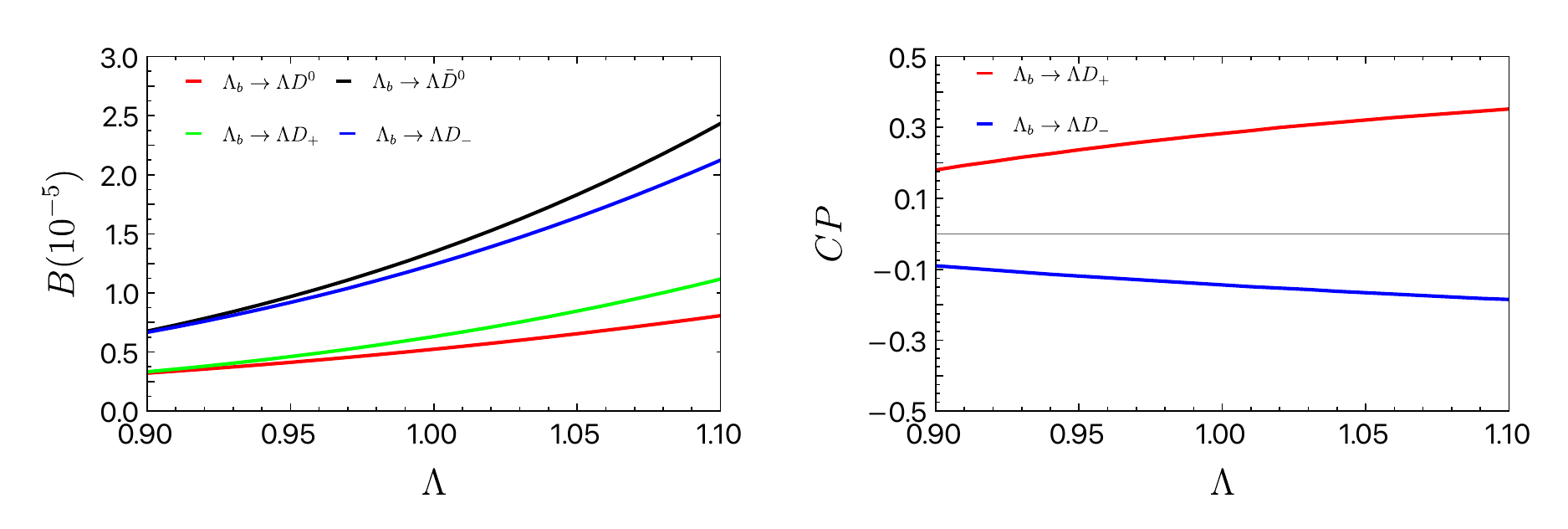}
		\caption{The dependence of the predicted branching fractions (left panel) and direct $C\!P$ asymmetries (right panel) on the cutoff parameter $\Lambda$ for the $\Lambda_b^0 \to \Lambda D$ decay channels.}
		\label{fig:Lambda_dependence}
	\end{figure}

	Building upon the established theoretical framework and input parameters, we calculate the total branching fractions as well as the $S$-wave and $P$-wave partial-wave branching fractions for each $\Lambda_b^0 \to \Lambda D$ decay channel, adopting a cutoff parameter of $\Lambda = 1.0 \pm 0.1$ GeV. These results are presented in Table~\ref{tab:Br_}. In addition, we compute the averaged decay asymmetry parameters $\langle \alpha \rangle$, $\langle \beta \rangle$, and $\langle \gamma \rangle$, alongside the corresponding $C\!P$ violation observables $a_{CP}^{\alpha}$, $a_{CP}^{\beta}$, and $a_{CP}^{\gamma}$ for the CP-eigenstate decays $\Lambda_b^0 \to \Lambda D_\pm$. These findings are summarized in Table~\ref{tab:cp_asymmetry_parameters} and~\ref{tab:decay_asymmetry_parameters}. To assess the sensitivity of our theoretical predictions to the phenomenological parameters, the dependence of the branching fractions and $C\!P$ asymmetries on the cutoff parameter $\Lambda$ is illustrated in Fig.~\ref{fig:Lambda_dependence}. 
	Furthermore, to clearly illustrate the dynamical impact of long-distance rescattering, Table~\ref{tab:br_comparison} compares our purely SD results and our comprehensive results incorporating final-state rescattering against existing theoretical predictions. These include a recent calculation within the PQCD framework~\cite{Rui:2026ihu}, the bag model (BM)~\cite{Geng:2022osc}, the phenomenological factorization approach (PFA)~\cite{Zhu:2018jet}, and the generalized factorization approach (GFA)~\cite{Giri:2001ju}.
	
	\begin{itemize}
		\item \textbf{Partial-Wave Branching Fractions and $S$-Wave Dominance:} 
		
		Table~\ref{tab:Br_} presents the partial-wave branching fractions for each decay channel. The numerical results reveal that the $S$-wave branching fraction is significantly larger than its $P$-wave counterpart across all channels, indicating that the decay yields are strongly dominated by the $S$-wave contributions. Specifically, the ratios are $\mathcal{B}^S(\Lambda_b^0 \to \Lambda D_+)/\mathcal{B}^P(\Lambda_b^0 \to \Lambda D_+) \approx 5.0$ and ${\mathcal{B}}^S(\Lambda_b^0 \to \Lambda D_-)/\mathcal{B}^P(\Lambda_b^0 \to \Lambda D_-) \approx 6.8$. Across the various decay channels, the relative magnitudes of the partial-wave contributions exhibit significant variation. Specifically, the $\Lambda_b^0 \to \Lambda \bar{D}^0$ decay demonstrates the most extreme $S$-wave dominance ($\mathcal{B}^S/\mathcal{B}^P \approx 11.3$), whereas the $\Lambda_b^0 \to \Lambda D^0$ mode displays the largest relative $P$-wave contribution ($\mathcal{B}^S/\mathcal{B}^P \approx 2.4$). These relative magnitudes dynamically determine the theoretical predictions for observable quantities, such as the decay asymmetry parameters.
		
		\item \textbf{Partial-Wave $C\!P$ Asymmetries and $P$-Wave Significance:}
		
		As demonstrated in Table~\ref{tab:cp_asymmetry_parameters}, although the $S$-wave dominates the branching fractions, the scenario is drastically different regarding $C\!P$ violation. For the $\Lambda D_+$ final state, we obtain $a_{CP}^S = 0.14$ and $a_{CP}^P = 0.98$ both are negative, with a relative magnitude of $|a_{CP}^P| / |a_{CP}^S| \approx 6.8$. For the $\Lambda D_-$ final state, the asymmetries are $a_{CP}^S = -0.07$ and $a_{CP}^P = -0.65$, yielding a ratio of $|a_{CP}^P| / |a_{CP}^S| \approx 9.5$. These findings underscore a critical physical mechanism: while the $S$-wave dictates the total decay width, the $P$-wave exhibits an extraordinary sensitivity to $C\!P$-violating phases.
		
		\item \textbf{Decay Asymmetry Parameter Analysis:}
		
		As shown in Table~\ref{tab:decay_asymmetry_parameters}, the decay asymmetry parameters $\alpha$ for both the $\Lambda D_+$ and $\Lambda D_-$ final states are negative. In particular, the absolute value for the $\Lambda D_+$ channel approaches unity, reflecting a highly pronounced interference effect. Notably, a recent PQCD calculation similarly predicts large negative values for the $\alpha$ parameter. However, significant numerical differences remain between the two theoretical approaches. This discrepancy fundamentally stems from their distinct theoretical frameworks. Specifically, the PQCD calculation relies primarily on short-distance non-factorizable contributions (such as internal $W$-emission and $W$-exchange diagrams). In contrast, our framework explicitly incorporates long-distance contributions via the final-state rescattering mechanism. It is precisely these long-distance effects that dominantly generate the strong phases essential for $C\!P$ violation.
		
		The feasibility of utilizing $\alpha$ to extract the CKM angle $\gamma$ has been systematically investigated in Ref.~\cite{Zhang:2021sit}. That study provided a theoretical range for $\alpha$ spanning from $-0.604$ to $0.389$. Our predicted value for the $\Lambda D_-$ final state ($-0.22$) falls safely within this range. Furthermore, it aligns closely with the fitted results of $-0.217$ (obtained with $r = 0.5$ and a phase combination of $0^\circ/150^\circ$) and $-0.277$ (obtained with $r = 2.0$ and $150^\circ/60^\circ$). In contrast, our predicted value for the $\Lambda D_+$ final state ($-0.85$) lies significantly outside this parameter space. This indicates that the $\Lambda D_+$ channel harbors much stronger dynamical interference effects than current phenomenological models suggest, presenting a highly promising new scenario for future experimental exploration.
		
		\item \textbf{Branching Fraction Comparison and Final-State Rescattering Effects:}
		
		Table~\ref{tab:br_comparison} highlights significant discrepancies among theoretical models. For the $\Lambda_b^0 \to \Lambda D^0(D_\pm)$ decays, our results are of the same order of magnitude as the predictions from the BM and GFA. However, for the highly suppressed $\Lambda_b^0 \to \Lambda \bar{D}^0$ mode, our predicted branching fraction of $(1.348_{-0.671}^{+1.084}) \times 10^{-5}$ exceeds those of other models by more than an order of magnitude. This dramatic enhancement originates directly from the rescattering mechanism, which effectively compensates for the highly suppressed short-distance amplitude via hadronic loops involving intermediate states such as $\Lambda_c^+ K^{(*)}$ and $p D_s^{(*)}$. In the purely SD calculation, the ratio $\mathcal{B}^{\text{SD}}(\Lambda_b^0 \to \Lambda \bar{D}^0) / \mathcal{B}^{\text{SD}}(\Lambda_b^0 \to \Lambda D^0) \approx 0.148$ is strictly consistent with the corresponding CKM suppression factor, $|(V_{ub}V_{cs}^*)/(V_{cb}V_{us}^*)|^2 \approx 0.142$. Remarkably, upon incorporating the LD contributions, the $\bar{D}^0$ channel is enhanced by approximately two orders of magnitude (a factor of 105), while the $D^0$ channel is enhanced merely by a factor of 6. Consequently, the $\bar{D}^0$ branching fraction surpasses that of the $D^0$ channel, yielding $\mathcal{B}(\Lambda_b^0 \to \Lambda \bar{D}^0) / \mathcal{B}(\Lambda_b^0 \to \Lambda D^0) \approx 2.6$. This striking numerical inversion provides a clear experimental criterion: observing $\mathcal{B}(\Lambda_b^0 \to \Lambda \bar{D}^0) > \mathcal{B}(\Lambda_b^0 \to \Lambda D^0)$ at the LHCb would unambiguously validate the indispensable role of LD rescattering dynamics.
		
		\item \textbf{Experimental Observable Channel Predictions:}
		
		Using the predicted branching fractions $\mathcal{B}(\Lambda_b^0 \to \Lambda D_\pm)$ and the precisely measured $D^0$ decay branching fractions ($\mathcal{B}(D^0 \to \pi^+\pi^-) = 1.454 \times 10^{-3}$, $\mathcal{B}(D^0 \to K^+K^-) = 4.08 \times 10^{-3}$, and $\mathcal{B}(D^0 \to K_S^0\pi^0) = 1.24 \times 10^{-2}$~\cite{ParticleDataGroup:2022pth}), we evaluate the branching fractions of the cascade decays:
		\begin{align}
			\mathcal{B}[\Lambda_b^0 \to \Lambda D_+(\to \pi^+\pi^-)] &\approx \left(0.917_{-0.433}^{+0.707}\right) \times 10^{-8}, \nonumber\\
			\mathcal{B}[\Lambda_b^0 \to \Lambda D_+(\to K^+K^-)] &\approx \left(2.574_{-1.216}^{+1.983}\right) \times 10^{-8}, \nonumber\\
			\mathcal{B}[\Lambda_b^0 \to \Lambda D_-(\to K_S^0\pi^0)] &\approx \left(1.538_{-0.712}^{+1.093}\right) \times 10^{-7}.
		\end{align}
		These theoretical predictions provide a solid benchmark for reconstructing $\Lambda_b^0$ decays via hadronic $D$ meson cascades, facilitating signal extraction and event yield estimations in current and future runs at the LHCb.
	\end{itemize}
	
	\begin{table}[htbp]
		\centering
		\caption{Branching fractions and partial-wave branching fractions of the $\Lambda_b^0 \to \Lambda D$ decays.}
		\label{tab:Br_}
		\renewcommand{\arraystretch}{1.3}
		\begin{tabular}{lccc}
			\hline\hline
			\textbf{Decay Channel} & $\mathcal{B}$ ($10^{-5}$) & $\mathcal{B}^S$ ($10^{-5}$) &  $\mathcal{B}^P$ ($10^{-5}$) \\
			\hline
			$\Lambda_b^0 \to \Lambda D^0$        & $0.524_{-0.202}^{+0.284}$ & $0.368_{-0.170}^{+0.244}$  & $0.155_{-0.032}^{+0.040}$  \\
			$\Lambda_b^0 \to \Lambda \bar{D}^0$  & $1.348_{-0.671}^{+1.084}$ & $1.238_{-0.611}^{+0.972}$  & $0.109_{-0.059}^{+0.113}$  \\
			$\Lambda_b^0 \to \Lambda D_+$        & $0.631_{-0.298}^{+0.486}$ & $0.525_{-0.259}^{+0.413}$  & $0.106_{-0.044}^{+0.078}$ \\
			$\Lambda_b^0 \to \Lambda D_-$      & $1.241_{-0.574}^{+0.882}$ & $1.082_{-0.523}^{+0.803}$  & $0.159_{-0.051}^{+0.100}$   \\
			\hline\hline
		\end{tabular}
	\end{table}
	
	\begin{table}[htbp]
		\centering
		\renewcommand{\arraystretch}{1.3}
		\setlength{\tabcolsep}{4pt}
		\caption{Comparison of theoretical predictions for the branching fractions of the $\Lambda_b^0 \to \Lambda D$ decays across different models. The abbreviations denote short-distance (SD) contributions, final-state rescattering, perturbative QCD (PQCD), the bag model (BM), the phenomenological factorization approach (PFA), and the generalized factorization approach (GFA).}
		\label{tab:br_comparison}
		\begin{tabular}{@{}lcccccc@{}}
			\hline\hline
			& \multicolumn{6}{c}{$\mathbf{\mathcal{B}} \ (10^{-5})$} \\
			\cline{2-7}
			\textbf{Decay Channel} & \textbf{This work (SD)} & \textbf{This work } & \textbf{PQCD}~\cite{Rui:2026ihu} & \textbf{BM}~\cite{Geng:2022osc} & \textbf{PFA}~\cite{Zhu:2018jet} & \makebox[1.2cm]{\textbf{GFA}}~\cite{Giri:2001ju} \\
			\hline
			$\Lambda_b^0 \to \Lambda D^0$       & $0.088$ & $0.524_{-0.202}^{+0.283}$ & $3.1_{-0.8}^{+1.8}$ & $0.66 \pm 0.06$ & $0.379$ & $0.456$ \\
			$\Lambda_b^0 \to \Lambda \bar{D}^0$ & $0.013$ & $1.348_{-0.671}^{+1.084}$ & $2.3_{-0.6}^{+1.1}$ & $0.09 \pm 0.01$ & $0.054$ & $0.083$ \\
			$\Lambda_b^0 \to \Lambda D_+$       & $0.036$ & $0.631_{-0.298}^{+0.486}$ & $1.9_{-0.4}^{+1.1}$& $0.29 \pm 0.03$ & ---     & ---       \\
			$\Lambda_b^0 \to \Lambda D_-$       & $0.064$ & $1.241_{-0.574}^{+0.882}$ &  $3.5_{-1.0}^{+1.9}$ & $0.47 \pm 0.05$ & ---     & ---    \\
			\hline\hline
		\end{tabular}
	\end{table}
	
	\begin{table*}[htbp]
		\centering
		\renewcommand{\arraystretch}{1.3}
		\caption{Comparison of $C\!P$ asymmetry parameters for the $\Lambda_b^0 \to \Lambda D_\pm$ decays between our work and the PQCD approach.}
		\label{tab:cp_asymmetry_parameters}
		\begin{tabular}{@{}lcccccc@{}}
			\hline\hline
			\textbf{Decay Channel} & $a_{CP}^S$ & $a_{CP}^P$ & $a_{CP}^{\text{dir}}$ & $a_{CP}^{\alpha}$ & $a_{CP}^{\beta}$ & $a_{CP}^{\gamma}$ \\
			\hline
			This work $\Lambda_b^0 \to \Lambda D_+$ & $0.14_{-0.13}^{+0.09}$ & $0.98_{-0.12}^{+0.01}$ & $0.28_{-0.11}^{+0.07}$ & $-0.47_{-0.07}^{+0.09}$ & $-0.05_{-0.10}^{+0.09}$ & $-0.25_{-0.02}^{+0.02}$\\
			PQCD~\cite{Rui:2026ihu}$\Lambda_b^0 \to \Lambda D_+$ & -- & -- & $0.72_{-0.14}^{+0.04}$ & $-0.11_{- 0.06}^{+0.06}$ & $-0.25_{-0.09}^{+0.11}$ & $-0.18_{-0.09}^{+0.06}$\\\hline
			This work $\Lambda_b^0 \to \Lambda D_-$ & $-0.07_{-0.04}^{+0.06}$ & $-0.65_{-0.10}^{+0.13}$ & $-0.14_{-0.04}^{+0.06}$ & $0.23_{-0.03}^{+0.05}$ & $-0.00_{-0.03}^{+0.03}$ & $0.13_{-0.00}^{+0.10}$  \\
			PQCD~\cite{Rui:2026ihu}$\Lambda_b^0 \to \Lambda D_-$ & -- & -- & $-0.44_{-0.03}^{+0.10}$ & $0.04^{+ 0.02}_{-0.01}$ & $0.12_{-0.05}^{+0.05}$ & $0.08_{-0.03}^{+0.05}$\\
			\hline\hline
		\end{tabular}
	\end{table*}
	
	\begin{table*}[htbp]
		\centering
		\renewcommand{\arraystretch}{1.3}
		\caption{Comparison of decay asymmetry parameters and their averages for the $\Lambda_b^0 \to \Lambda D_\pm$ decays between our work and the PQCD approach.}
		\label{tab:decay_asymmetry_parameters}
		\begin{tabular}{@{}lcccccc@{}}
			\hline\hline
			\textbf{Decay Channel} & $\alpha$ & $\beta$ & $\gamma$ & $\langle\alpha\rangle$ & $\langle\beta\rangle$ & $\langle\gamma\rangle$ \\
			\hline
			This work $\Lambda_b^0 \to \Lambda D_+$ & $-0.85_{-0.03}^{+0.05}$ & $-0.22_{-0.01}^{+0.02}$ & $0.48_{-0.11}^{+0.05}$ & $-0.37_{-0.04}^{+0.04}$ & $-0.16_{-0.09}^{+0.08}$ & $0.74_{-0.08}^{+0.03}$\\
			PQCD~\cite{Rui:2026ihu}$\Lambda_b^0 \to \Lambda D_+$ & $-0.96_{-0.03}^{+0.02}$ & $-0.16^{+0.07}_{-0.08}$ & $0.22_{-0.13}^{+0.06}$ & -- & -- & --\\
			\hline
			This work $\Lambda_b^0 \to \Lambda D_-$ & $-0.22_{-0.04}^{+0.07}$ & $-0.39_{-0.15}^{+0.13}$ & $0.90_{-0.07}^{+0.04}$ & $-0.45_{-0.01}^{+0.02}$ & $-0.38_{-0.11}^{+0.08}$ & $0.76_{-0.07}^{+0.04}$  \\
			PQCD~\cite{Rui:2026ihu}$\Lambda_b^0 \to \Lambda D_-$ & $-0.90_{-0.05}^{+0.05}$ & $0.15^{+0.08}_{-0.11}$ & $0.40^{+0.09}_{-0.11}$ & -- & -- & -- \\
			\hline\hline
		\end{tabular}
	\end{table*}
	\section{Conclusion}
	\label{sec:conclusion}
	
	In summary, we have performed a systematic investigation of the non-leptonic $\Lambda_b^0 \to \Lambda D$ decays by evaluating both the short-distance contributions under the factorization hypothesis and the long-distance non-factorizable contributions via the final-state rescattering mechanism. Our findings underscore the indispensable role of long-distance strong interactions in achieving a comprehensive understanding of bottom-baryon decay dynamics. We reveal a pronounced dynamical disparity within the partial-wave sectors: while the $S$-wave amplitude strictly dominates the total branching fractions across all channels, the $P$-wave component exhibits a significantly magnified $C\!P$ asymmetry. The predicted non-zero direct $C\!P$ asymmetries demonstrate that final-state rescattering effectively provides the essential strong phases, which are typically highly suppressed in pure short-distance calculations. Furthermore, our derived decay asymmetry parameter $\alpha$ for the $\Lambda D_-$ final state falls well within the preferred parameter space required for the precise extraction of the CKM angle $\gamma$~\cite{Zhang:2021sit}.
	
	Crucially, our theoretical framework resolves the notable shortcomings of previous phenomenological models in evaluating color-suppressed decay channels. The incorporation of rescattering effects introduces a substantial enhancement to the $\Lambda_b^0 \to \Lambda \bar{D}^0$ decay amplitude, culminating in the unequivocal prediction that its branching fraction will exceed that of the $\Lambda D^0$ channel. This distinctive hierarchy inversion provides a robust and testable experimental signature to explicitly validate the rescattering mechanism. Finally, the predicted observable branching fractions for the $D_+$ and $D_-$ $C\!P$ eigenstates via their hadronic cascade decays establish a direct theoretical benchmark for future experimental measurements. Ultimately, these results offer promising theoretical pathways to reliably test the SM and probe for potential new physics signals in the heavy baryon sector.
	\appendix
	\section{Effective Lagrangians and Strong Coupling Constants}
	\label{sec:appendix_A}
	
	Within the framework of $SU(3)$ flavor symmetry, the effective Lagrangians describing the strong interactions for the relevant hadronic vertices $\mathcal{B}_c\mathcal{B}P$, $\mathcal{B}_c\mathcal{B}V$, $D^*DP$, $D^*D^*P$, $DDV$, $D^*DV$, and $D^*D^*V$ are constructed as follows:
	\begin{align}
		\mathcal{L}_{\mathcal{B}_c \mathcal{B} P} &= g_{P \mathcal{B}_c \mathcal{B}} \bar{\mathcal{B}}^i_c i\gamma_{5} P_j^k \mathcal{B}_k^j, \label{eq:L_BBP} \\
		\mathcal{L}_{\mathcal{B}_c \mathcal{B} V} &= g_{V \mathcal{B}_c \mathcal{B}} \bar{\mathcal{B}}^i_c \gamma_{\mu} V_j^{\mu k} \mathcal{B}^j_k + \frac{f_{V \mathcal{B}_c\mathcal{B}}}{m_{\mathcal{B}_c}+m_{\mathcal{B}}} \bar{\mathcal{B}}^i_c \sigma_{\mu\nu} \partial^{\mu} V_j^{\nu k} \mathcal{B}^j_k, \label{eq:L_BBV} \\
		\mathcal{L}_{D^* DP} &= -i g_{D^* D P} \left( D_i \partial^\mu P^i_j D_\mu^{* j \dagger} - D_{i \mu}^* \partial^\mu P^i_j D^{j \dagger} \right), \label{eq:L_DsDP} \\
		\mathcal{L}_{D^* D^* P} &= \frac{1}{2} g_{D^* D^* P} \varepsilon_{\mu \nu \alpha \beta} D_i^{* \mu} \partial^\nu P^i_j \overleftrightarrow{\partial}^\alpha D^{* \beta j \dagger}, \label{eq:L_DsDsP} \\
		\mathcal{L}_{DDV} &= -i g_{D D V} D_i \overleftrightarrow{\partial}_\mu D^{j\dagger} \left(V^\mu\right)_j^i, \label{eq:L_DDV} \\
		\mathcal{L}_{D^* DV} &= -2 g_{D^* D V} \varepsilon_{\mu \nu \alpha \beta} \left(\partial^\mu V^\nu\right)_j^i \left( D_i \overleftrightarrow{\partial}^\alpha D^{* \beta j\dagger} - D_i^{* \beta} \overleftrightarrow{\partial}^\alpha D^{j\dagger} \right), \label{eq:L_DsDV} \\
		\mathcal{L}_{D^* D^* V} &= i g_{D^* D^* V} D_i^{* \nu} \overleftrightarrow{\partial}_\mu D_\nu^{* j\dagger} \left(V^\mu\right)_j^i + 4 i f_{D^* D^* V} D_{i \mu}^* \left(\partial^\mu V^\nu - \partial^\nu V^\mu\right)_j^i D_\nu^{* j\dagger}, \label{eq:L_DsDsV}
	\end{align}
	where $i, j, k$ denote the $SU(3)$ flavor indices. Here, $\mathcal{B}_c$ and $\mathcal{B}$ represent the charmed baryon anti-triplet and the light baryon octet, respectively. The fields $P$ ($V$) correspond to the light pseudoscalar (vector) meson nonets, while $D_i$ ($D_i^*$) denote the charmed pseudoscalar (vector) meson anti-triplets, given by $D_i = (D^0, D^+, D_s^+)$.
	
	The strong coupling constants for these vertices are systematically derived based on $SU(3)$ flavor symmetry relations. Their explicit values are summarized in Table~\ref{tab:unified_coupling_constants} and Table~\ref{tab:meson_coupling_constants}, with the fundamental benchmark parameters taken from Refs.~\cite{Cheng:2004ru,Aliev:2006xr,Aliev:2009ei,Yu:2017zst,Duan:2024zjv}.
	
	\begin{table}[htbp]
		\centering
		\renewcommand{\arraystretch}{1.5}
		\caption{Strong coupling constants for the $\mathcal{B}_c \mathcal{B} P$ and $\mathcal{B}_c \mathcal{B} V$ vertices derived from $SU(3)$ flavor symmetry. The numerical parameters are fixed as: $g_{\Lambda_c^+ p D^0} = a_1 = 4.82$, $g_{\Lambda_c^+ p D^{*0}} = b_1 = 2.05$, and $f_{\Lambda_c^+ p D^{*0}} = b^{\prime}_1 = 7.78$.}
		\label{tab:unified_coupling_constants}
		\begin{tabular}{@{}lclclclc@{}}
			\hline\hline
			\multicolumn{8}{c}{\textbf{Pseudoscalar Meson Vertices ($\mathcal{B}_c \mathcal{B} P$)}} \\
			\hline
			\textbf{Vertex} & $g$ & \textbf{Vertex} & $g$ & \textbf{Vertex} & $g$ & \textbf{Vertex} & $g$ \\
			\hline
			$\Lambda_c^+ \to \Lambda D_s^+$ & $-\sqrt{\frac{2}{3}} a_1$ & $\Lambda_c^+ \to p D^0$ & $a_1$ & $\Lambda_c^+ \to n D^+$ & $a_1$ & $\Xi_c^+ \to \Lambda D^+$ & $-\frac{a_1}{\sqrt{6}}$ \\
			$\Xi_c^+ \to \Sigma^0 D^+$ & $\frac{a_1}{\sqrt{2}}$ & $\Xi_c^+ \to \Sigma^+ D^0$ & $-a_1$ & $\Xi_c^+ \to \Xi^0 D_s^+$ & $-a_1$ & $\Xi_c^0 \to \Lambda D^0$ & $\frac{a_1}{\sqrt{6}}$ \\
			$\Xi_c^0 \to \Sigma^- D^+$ & $a_1$ & $\Xi_c^0 \to \Sigma^0 D^0$ & $\frac{a_1}{\sqrt{2}}$ & $\Xi_c^0 \to \Xi^- D_s^+$ & $a_1$ & & \\
			\hline
			\multicolumn{8}{c}{\textbf{Vector Meson Vertices ($\mathcal{B}_c \mathcal{B} V$)}} \\
			\hline
			\textbf{Vertex} & $g$ & \textbf{Vertex} & $f$ & \textbf{Vertex} & $g$ & \textbf{Vertex} & $f$ \\
			\hline
			$\Lambda_c^+ \to \Lambda D_s^{*+}$ & $-\sqrt{\frac{2}{3}} b_1$ & $\Lambda_c^+ \to \Lambda D_s^{*+}$ & $-\sqrt{\frac{2}{3}} b^{\prime}_1$ & $\Lambda_c^+\to p D^{*0} $ & $ b_1$ & $\Lambda_c^+\to p D^{*0} $ & $ b^{\prime}_1$  \\
			$\Lambda_c^+\to n D^{*+} $ & $ b_1$ & $\Lambda_c^+\to n D^{*+} $ & $ b^{\prime}_1$ & $\Xi_c^+\to \Lambda D^{*+} $ & $ -\frac{b_1}{\sqrt{6}}$ & $\Xi_c^+\to \Lambda D^{*+} $ & $ -\frac{b^{\prime}_1}{\sqrt{6}}$ \\
			$\Xi_c^+\to \Sigma^0 D^{*+} $ & $ \frac{b_1}{\sqrt{2}}$ & $\Xi_c^+\to \Sigma^0 D^{*+} $ & $ \frac{b^{\prime}_1}{\sqrt{2}}$ & $\Xi_c^+\to \Sigma^+ D^{*0} $ & $ -b_1$ & $\Xi_c^+\to \Sigma^+ D^{*0} $ & $ -b^{\prime}_1$\\
			$\Xi_c^+\to \Xi^0 D^{*+}_s $ & $ -b_1$ & $\Xi_c^+\to \Xi^0 D^{*+}_s $ & $ -b^{\prime}_1$ & $\Xi_c^0\to \Lambda D^{*0} $ & $ \frac{b_1}{\sqrt{6}}$ & $\Xi_c^0\to \Lambda D^{*0} $ & $ \frac{b^{\prime}_1}{\sqrt{6}}$\\
			$\Xi_c^0\to \Sigma^- D^{*+} $ & $ b_1$ & $\Xi_c^0\to \Sigma^- D^{*+} $ & $ b^{\prime}_1$ & $\Xi_c^0\to \Sigma^0 D^{*0} $ & $ \frac{b_1}{\sqrt{2}}$ & $\Xi_c^0\to \Sigma^0 D^{*0} $ & $ \frac{b^{\prime}_1}{\sqrt{2}}$\\
			$\Xi_c^0\to \Xi^- D^{*+}_s $ & $ b_1$ & $\Xi_c^0\to \Xi^- D^{*+}_s $ & $ b^{\prime}_1$ & & & & \\
			\hline\hline
		\end{tabular}
	\end{table}
	
	\begin{table}[htbp]
		\centering
		\renewcommand{\arraystretch}{1.5}
		\caption{Strong coupling constants for the $D^*DV$, $D^*DP$, and $DDV$ vertices. These constants are derived from the benchmark couplings $g_{D^*D\rho} = d_1 = 2.345$, $g_{D^*D\pi} = e_1 = 17.9$, and $g_{DD\rho} = f_1 = 3.69$, combined with $SU(3)$ flavor symmetry relations.}
		\label{tab:meson_coupling_constants}
		\begin{tabular}{@{}lclclclc@{}}
			\hline\hline
			\textbf{Vertex} & $g$ & \textbf{Vertex} & $g$ & \textbf{Vertex} & $g$  & \textbf{Vertex} & $g$\\
			\hline
			$D^{*0} \to D^0 \rho^0$ & $\frac{d_1}{\sqrt{2}}$ & $D^{*0} \to D^+ \rho^-$ & $d_1$ & $D^{*0} \to D_s^+ K^{*-}$ & $d_1$ &$D^{*+} \to D^0 \rho^+$ & $d_1$ \\ 
			$D^{*+} \to D^+ \rho^0$ & $-\frac{d_1}{\sqrt{2}}$& $D^{*+} \to D_s^+ \bar{K}^{*0}$ & $d_1$ & $D_s^{*+} \to D^0 K^{*+}$ & $d_1$ & $D_s^{*+} \to D^+ K^{*0}$ & $d_1$ \\ 
			$D^{*0} \to D^0 \pi^0$ & $\frac{e_1}{\sqrt{2}}$ & $D^{*0} \to D^+ \pi^-$ & $e_1$ & $D^{*0} \to D_s^+ K^-$ & $e_1$ & $D^{*+} \to D^0 \pi^+$ & $e_1$ \\
			$D^{*+} \to D^+ \pi^0$ & $-\frac{e_1}{\sqrt{2}}$ & $D^{*+} \to D_s^+ \bar{K}^0$ & $e_1$ & $D_s^{*+} \to D^0 K^+$ & $e_1$ & $D_s^{*+} \to D^+ K^0$ & $e_1$ \\ 
			$D_s^+ \to D^0 K^{*+}$ & $f_1$ & $D_s^+ \to D^+ K^{*0}$ & $f_1$ & $D^0 \to D^0 \rho^0$ & $\frac{f_1}{\sqrt{2}}$ & $D^0 \to D^+ \rho^-$ & $f_1$ \\ 
			$D^0 \to D_s^+ K^{*-}$ & $f_1$ & $D^+ \to D^0 \rho^+$ & $f_1$ & $D^+ \to D^+ \rho^0$ & $-\frac{f_1}{\sqrt{2}}$ & $D^+ \to D_s^+ \bar{K}^{*0}$ & $f_1$ \\
			\hline\hline
		\end{tabular}
	\end{table}
	For the $\Lambda_b^0 \to \Lambda \bar{D}^0$ decay channel, the integrated amplitudes corresponding to the triangle diagrams in Fig.~\ref{fig:LambdaD}(a-c) are explicitly given by:
	\begin{align}
		\mathcal{M}_{a}[D_s^{*-},p;K^-] &= g_{D^{*-}_s D^0 K^{-}} g_{p \Lambda K^- } \int \frac{d^4k}{(2\pi)^4} \bar{u}(p_4,\lambda_4)\gamma_5 (\not\!p_2+m_2)(B_1\gamma_{\nu}\gamma_5 + B_2\frac{p_{2\nu}}{m_{\Lambda^0_b}}\gamma_5 . \nonumber \\
		&\quad  + B_3\gamma_{\nu} + B_4\frac{p_{2\nu}}{m_{\Lambda^0_b}}) u(p_i,\lambda_i)\frac{-(-g^{\alpha\nu}+\frac{p_1^{\alpha} p_1^{\nu}}{m_1^2})k_{\alpha}{\mathcal{F}}}{(p_1^2-m_1^2+i\varepsilon)(p_2^2-m_2^2+i\varepsilon)(k^2-m_k^2+i\varepsilon)}, \nonumber\\
		\mathcal{M}_{a}[D_s^{*-},\Lambda^+_c;K^{*-}] &= 2ig_{D^{*-}_s D^0 K^{*-}} \int \frac{d^4k}{(2\pi)^4}\bar{u}(p_4,\lambda_4)(-g_{p \Lambda K^{*-}}\gamma^{\mu} - \frac{if_{p \Lambda K^{*-}}}{m_2+m_4}\sigma^{\mu\nu}k_{\nu}) k^{\alpha} \nonumber \\
		&\quad \times (\not\!p_2 +m_2)(B_1\gamma_{\delta}\gamma_5 + B_2\frac{p_{2\delta}}{m_i}\gamma_5 + B_3\gamma_{\delta} + B_4\frac{p_{2\delta}}{m_i}) u(p_i,\lambda_i)\varepsilon_{\alpha\beta\rho\sigma}(p_3+p_1)^{\rho} \nonumber \\
		&\quad \times \frac{(-g^{\delta\beta}+\frac{p_1^{\delta} p_1^{\beta}}{m_1^2})(-g^{\sigma\mu}+\frac{k^{\sigma} k^{\mu}}{m_k^2})\mathcal{F}}{(p_1^2-m_1^2+i\varepsilon)(p_2^2-m_2^2+i\varepsilon)(k^2-m_k^2+i\varepsilon)},  \nonumber\\
		\mathcal{M}_{a}[D_s^-,p;K^{*-}] &= g_{D^-_s D^0 K^{*-}}\int\frac{d^4k}{(2\pi)^4}\bar{u}(p_4,\lambda_4)(-g_{p \Lambda K^{*-}}\gamma^{\mu} - \frac{if_{p \Lambda K^{*-}}}{m_2+m_4}\sigma^{\mu\nu}k_{\nu})(\not\!p_2+m_2) \nonumber \\
		&\quad \times \frac{(A_1+A_2\gamma_5)u(p_i,\lambda_i)\left(-g_{\mu\alpha}+\frac{k_\mu k_\alpha}{m_k^2}\right)(p_1+p_3)_{\alpha}{\mathcal{F}}}{(p_1^2-m_1^2+i\varepsilon)(p_2^2-m_2^2+i\varepsilon)(k^2-m_k^2+i\varepsilon)},  \nonumber\\
		\mathcal{M}_{b}[D_s^-,p;\Lambda^+_c] &= \int \frac{d^4k}{(2\pi)^4} \bar{u}(p_4,\lambda_4) g_{\Lambda \Lambda_c^+ D_s^-} g_{\Lambda_c^+ p D^0} \gamma_5(\not\!k+m_k)\gamma_5(\not\!p_2+m_2) \nonumber \\
		&\quad \times \frac{(A_1+A_2\gamma_5)u(p_i,\lambda_i) \mathcal{F}}{(p_1^2-m_1^2+i\varepsilon)(p_2^2-m_2^2+i\varepsilon)(k^2-m_k^2+i\varepsilon)},  \nonumber\\
		\mathcal{M}_{b}[D_s^{*-},\Lambda^+_c;p] &= \int\frac{d^{4}k}{(2\pi)^{4}} g_{\Lambda_c^+ p D^0}\bar{u}(p_4,\lambda_4)(-g_{\Lambda \Lambda_c^+ D_s^{*-}}\gamma_{\mu} - \frac{if_{\Lambda \Lambda_c^+ D_s^{*-}}}{m_2+m_4}\sigma_{\mu\nu}p_1^{\nu}\nonumber \\
		&\quad \times(\not\!k+m_k)\gamma_5  (\not\!p_2+m_2)(B_1\gamma_{\beta}\gamma_5 + B_2\frac{p_{2\beta}}{m_i}\gamma_5 + B_3\gamma_{\beta} + B_4\frac{p_{2\beta}}{m_i}) \nonumber \\
		&\quad \times(-g^{\beta \mu}+\frac{p_1^{\beta}p_1^\mu}{m_{1}^2})\frac{\mathcal{F}}{(p_1^2-m_1^2+i\varepsilon)(p_2^2-m_2^2+i\varepsilon)(k^2-m_k^2+i\varepsilon)},  \nonumber\\
		\mathcal{M}_{c}[D^0,\Lambda;\Xi^0_c] &= \int \frac{d^4k}{(2\pi)^4} \bar{u}(p_4,\lambda_4) g_{\Lambda \Xi_c^0 D^0} g_{\Xi_c^0 \Lambda D^0} \gamma_5(\not\!k+m_k)\gamma_5(\not\!p_2+m_2) u(p_i,\lambda_i)
		\nonumber \\
		&\quad \times \frac{(A_1+A_2\gamma_5)u(p_i,\lambda_i) \mathcal{F}}{(p_1^2-m_1^2+i\varepsilon)(p_2^2-m_2^2+i\varepsilon)(k^2-m_k^2+i\varepsilon)},  \nonumber\\
		\mathcal{M}_{c}[D^{*0},\Lambda;\Xi^0_c] &= \int\frac{d^{4}k}{(2\pi)^{4}} g_{\Lambda \Xi_c^0 D^0}\bar{u}(p_4,\lambda_4)(-g_{\Lambda \Xi_c^0 D^{*0}}\gamma_{\mu} - \frac{if_{\Lambda \Xi_c^0 D^{*0}}}{m_2+m_4}\sigma_{\mu\nu}p_1^{\nu})(\not\!k+m_k)\gamma_5 \nonumber \\
		&\quad \times (\not\!p_2+m_2)(B_1\gamma_{\beta}\gamma_5 + B_2\frac{p_{2\beta}}{m_i}\gamma_5 + B_3\gamma_{\beta} + B_4\frac{p_{2\beta}}{m_i})(-g^{\beta \mu}+\frac{p_1^{\beta}p_1^\mu}{m_{1}^2}) \nonumber \\
		&\quad \times \frac{\mathcal{F}}{(p_1^2-m_1^2+i\varepsilon)(p_2^2-m_2^2+i\varepsilon)(k^2-m_k^2+i\varepsilon)}u(p_i,\lambda_i).
	\end{align}
	
\end{document}